\documentstyle[12pt,amssymbols]{article}

\def\journal{\topmargin .3in    \oddsidemargin .5in
        \headheight 0pt \headsep 0pt
        \textwidth 5.625in 
\textheight 8.25in 
        \marginparwidth 1.5in
        \parindent 2em
        \parskip .5ex plus .1ex         \jot = 1.5ex}
\journal

\newskip\humongous \humongous=0pt plus 1000pt minus 1000pt

\newif\ifdtup

\begin{document}
\begin{titlepage}
\begin{center}
\today \hfill    LBL- 36334\\
          \hfill    FERMILAB-PUB-94/402-T\\

\vskip .5in

{\large \bf $W^+Z$ and $W^+\gamma^*$ Backgrounds to Strong $W^+W^+$ Scattering
at the LHC}\footnote
{This work was supported by the Director, Office of Energy
Research, Office of High Energy and Nuclear Physics, Division of High
Energy Physics of the U.S. Department of Energy under Contracts
DE-AC03-76SF00098 and DE-AC02-76CHO3000.}

\vskip .35in

Michael S. Chanowitz
 \\[.20in]

{\em Theoretical Physics Group\\
     Lawrence Berkeley Laboratory\\
     University of California\\
     Berkeley, California 94720}
 \\[.25in]
William B. Kilgore
 \\[.20in]
{\em Theoretical Physics Department\\
Fermi National Accelerator Laboratory\\
P.O. Box 500, Batavia, Illinois 60510}
\end{center}

\vskip .25in

\begin{abstract}
Because of inevitable blind spots at high rapidity and low transverse
momentum, the process $\overline qq \rightarrow l^+ l^- \ l^{+} \nu$
provides a surprisingly large background to the $W^+W^+
\rightarrow l^+ \nu \ l^+ \nu$ signal associated with strong $WW$ scattering.
Previous calculations of
the total background are approximately doubled, and the estimate of
the ``no-lose''
luminosity at the LHC to observe
manifestations of the electroweak symmetry breaking mechanism
is increased to between $\sim 100$ and
$\sim 140$ fb$^{-1}$.
\end{abstract}

\end{titlepage}
\renewcommand{\thepage}{\roman{page}}
\setcounter{page}{2}
\mbox{ }

\vskip 1in

\begin{center}
{\bf Disclaimer}
\end{center}

\vskip .2in

\begin{scriptsize}
\begin{quotation}
This document was prepared as an account of work sponsored by the United
States Government.  Neither the United States Government nor any agency
thereof, nor The Regents of the University of California, nor any of their
employees, makes any warranty, express or implied, or assumes any legal
liability or responsibility for the accuracy, completeness, or usefulness
of any information, apparatus, product, or process disclosed, or represents
that its use would not infringe privately owned rights.  Reference herein
to any specific commercial products process, or service by its trade name,
trademark, manufacturer, or otherwise, does not necessarily constitute or
imply its endorsement, recommendation, or favoring by the United States
Government or any agency thereof, or The Regents of the University of
California.  The views and opinions of authors expressed herein do not
necessarily state or reflect those of the United States Government or any
agency thereof of The Regents of the University of California and shall
not be used for advertising or product endorsement purposes.
\end{quotation}
\end{scriptsize}

\vskip 2in

\begin{center}
\begin{small}
{\it Lawrence Berkeley Laboratory is an equal opportunity employer.}
\end{small}
\end{center}

\newpage
\renewcommand{\thepage}{\arabic{page}}
\setcounter{page}{1}
\noindent {\it \underline {Introduction} }

In the search for the mechanism of electroweak symmetry breaking,
strong $WW$ scattering above 1 TeV in the $WW$ center of mass energy
is the  signal of ``last resort''  since it occurs
if other, typically larger signals below 1 TeV, such as Higgs bosons, do
not.\cite{strongww}
A collider with enough energy and luminosity to
observe strong $WW$ scattering can
provide evidence of the symmetry breaking mechanism whatever form it may
take. Using a QCD-like
chiral Lagrangian with a dominant ``$\rho$'' meson to study strong scattering
in
the $WZ$ and $W^+W^+$ channels,
we found a complementary relationship between
the resonant ``$\rho$'' signal, best observed in the $WZ$ channel, and
nonresonant scattering in the $W^+W^+$ channel.{\cite {mcwk1}} In that model
we found that
the worst case is for $m_{\rho} \simeq 2.5$ TeV, corresponding to minimal
one-doublet technicolor with $N_{TC} = 2$, for which the ``$\rho$'' is too
heavy to observe directly at the LHC
as a $WZ$ resonance but is still light enough for chiral dynamics associated
with $t$ and $u$-channel ``$\rho$'' exchange to
suppress nonresonant $W^+W^+$ scattering.
For experimental cuts that optimize
the signal relative to the ``irreducible'' backgrounds, we used this
worst case to estimate a ``no-lose'' luminosity
$\simeq 60$ fb$^{-1}$ for the LHC operating at 14 TeV. Similar
results\footnote{One
important difference is that we ``customize'' the experimental
cuts to obtain the optimal significance for each model while Bagger et al. use
a single set of cuts, optimized for the standard model with $m_H = 1$ TeV, for
all models.}
have been reported by Bagger et al.{\cite {baggeretal}}

The irreducible backgrounds are those with the same final state as the
signal. For strong $W^+W^+$ scattering the leading irreducible
backgrounds are the order
$\alpha_W^2$ and $\alpha_W \alpha_S$ amplitudes for $qq \rightarrow qqW^+W^+$
with the former computed in the standard model with a light Higgs boson,
$m_H \leq 100$ GeV. We consider leptonic decays, $W^+W^+ \rightarrow l^+ \nu
l^+ \nu$ where $l =$ $e$ or $\mu$.
In our previous study we used cuts on the $e^+$ and $\mu^+$ leptons and a
veto of events with high
$p_T$ jets in the central region, to enhance the
longitudinally polarized $W^+W^+$ pairs of the signal over the
transverse-transverse and transverse-longitudinal
pairs of the background.

In the $W^+W^+$ channel both we and the authors of reference {\cite
{baggeretal}} did
not use a tag on forward jets, seeking instead to isolate a clean
signal by means of harder cuts on the lepton variables. A jet tag may be
necessary if we hope to detect gauge boson pairs in ``mixed decays'',
where one boson decays to
hadrons and the other to leptons, but it may not be needed if both bosons
decay to leptons, as is essential for the like-sign $WW$ signal.
The use of hard leptonic cuts for purely leptonic decays
is especially effective for strong scattering models since they are
characterized by the hardest diboson energy distributions consistent with
unitarity. If this strategy
suffices it has the advantage of being cleaner theoretically and
experimentally.
Estimates of jet tagging efficiency are uncertain because
the transverse momentum and rapidity distributions of the forward jets
are probably sensitive to QCD corrections and because the jet
detection efficiency in the forward region is sensitive both to details of the
detector and to how the jets hadronize. If both strategies can be made to work
they will complement one another.

Other backgrounds (not irreducible) that can fake the $\l^+ l^+$ signal, from
$\overline tt${\cite {ttbar}} and $\overline ttW^+${\cite {ttbarW}}, can
be suppressed.
We have independently confirmed the conclusions of {\cite {ttbar}} for the
$\overline tt$ background. After cuts the $t$ quark backgrounds are much
smaller than the irreducible background from $qq \rightarrow qqW^+W^+$.

In a different approach to the $qq \rightarrow qqW^+W^+$ signal
Azuelos, Leroy, and Tafirout considered
tagging the two forward quark jets.{\cite {azuelosetal}} Their study
includes  a simplified simulation of the proposed ATLAS detector.
In addition to the irreducible backgrounds and the $W\overline tt$ background,
they also considered $\overline qq \rightarrow
W^+Z \rightarrow l^+ \nu \l^+ \l^-$ where
the $l^-$ escapes detection. The $\overline qq$ annihilation amplitude is
computed with PYTHIA, which includes gluon radiation.
To satisfy the jet tag two gluons must accompany the $WZ$ bosons, so that
the contribution of these $WZ$ events to the background is well
below that of the order $\alpha_W^2$ $qq \rightarrow qqW^+W^+$
amplitude, which is also the dominant background in their analysis.
It remains to evaluate the order $\alpha_W^2$ amplitude background from
$qq \rightarrow qqWZ$, which after the jet tags and central jet veto is
probably larger than $\overline qq \rightarrow WZ$.

We are motivated by the study of Azuelos et al.
to consider the effect of the $WZ$ background in our
approach. We have gone beyond the $WZ$ background
to include the full set of 10 tree level Feynman diagrams\footnote{The
amplitudes are evaluated using HELAS\cite{helas} as implemented in
MADGRAPH\cite{madgraph}.}
for the process $\overline qq \rightarrow l^{\prime +} \nu^{\prime} l^+ l^-$
where $l$ and
$l^{\prime}$ are electrons or muons.\footnote{We neglect the small interference
between the like-sign leptons when $l = l^{\prime}$.}
Three diagrams correspond to $WZ$ production and  three to $W\gamma^*$.
We find after cuts that the contributions
not attributable to $\overline qq \rightarrow
WZ$ constitute nearly half of the total
$\overline qq \rightarrow l^{\prime +} \nu^{\prime} l^+ l^-$ background.

After reviewing the model and the irreducible backgrounds
we describe the $\overline qq \rightarrow
\overline l\nu \overline ll$ background
and evaluate its effect on the observability of the $W^+W^+ + W^-W^-$
signal. We then briefly discuss
the sensitivity of the results to collider energy and detector
coverage and the background from $qq \rightarrow qqWZ$. We conclude
with a few remarks, including the observation that {\em strong scattering} in
$qq \rightarrow qqWZ$ provides a useful ``pseudo-amplification'' of the
strong $W^+W^+$ scattering signal.

\noindent {\it \underline {Signal and Irreducible Background}}

The computation of the signal and irreducible background is as in reference
\cite{mcwk1}, which the reader can consult for additional details and
references. Here we only sketch the essential points.

In \cite{mcwk1} we considered the strong scattering of longitudinally polarized
gauge bosons in the channels $qq
\rightarrow qqW^{\pm}Z$ and $qq \rightarrow qqW^+W^+/W^-W^-$.
If the electroweak $SU(2)_L \times U(1)_Y$ is broken by a strong force, the
scattering of longitudinally polarized gauge bosons of energy $E \gg M_W$
is approximately equal to that of the
corresponding unphysical Goldstone bosons.
At low energies compared to the mass scale of the new strong force
the Goldstone boson interactions
can be described by an effective Lagrangian like the chiral
Lagrangian that describes QCD. To explore the relationship between resonant
$WZ$ and nonresonant $W^+W^+$ strong scattering signals
we assumed QCD-like dynamics, with an
effective Lagrangian incorporating a dominant ``$\rho$'' meson\cite{rhomodel}
and with K-matrix
unitarization. Applied to QCD the model
provides a good description of both $\pi^+ \pi^0$ and
$\pi^+ \pi^+$ scattering data to unexpectedly high energy $\simeq 1.2$ GeV.

Applied to electroweak symmetry breaking the model exhibits
a complementary relationship between the resonant
$WZ$ channel and the nonresonant $W^+W^+$ channel.
For smaller $m_{\rho}$ the resonant $\rho \rightarrow WZ$ signal is large while
the nonresonant $W^+W^+$ signal is suppressed. For very large $m_{\rho}$ the
resonant $WZ$ signal is unobservable but nonresonant $W^+W^+$ scattering is
large, approaching the K-matrix unitarization of the low energy theorem as
$m_{\rho} \rightarrow \infty$.

In \cite{mcwk1} we considered only irreducible backgrounds to the
$W^+W^+$ signal. We used cuts
on the $l^+l^+$ decay products (rapidity, $\eta(l^+)$, transverse
momentum, $p_{T}(l^+)$, and  the azimuthal angle between the two
leptons, $\phi(l^+l^+)$)
that exploit the differing energy dependence and
polarization of signal and background. We also imposed
a veto on events with a central,
high $p_T$ jet, that also exploits the boson polarizations.
We optimized the cuts
in $p_T(l^+)$ and $\phi(l^+l^+)$ for each model
to establish the minimum luminosity for a significant signal.

The observability criterion in \cite{mcwk1} and in this paper is
\begin{equation}
\sigma^{\uparrow}   =  S/\sqrt{B}  \ge  5
\end{equation}
\begin{equation}
\sigma^{\downarrow}   =  S/\sqrt{S+B}  \ge  3
\end{equation}
\begin{equation}
S \ge B,
\end{equation}
where $S$ and $B$ are the number of signal and background events, and
$\sigma^{\uparrow}$ and $\sigma^{\downarrow}$ are respectively the number of
standard deviations for the background to fluctuate
up to give a false signal or for the signal plus background to
fluctuate down to the level of the background alone.
In addition we require $S \ge B$ so that the signal is unambiguous despite the
systematic uncertainty in the size of the backgrounds,
expected to be known to within $ \leq \pm 30 \%$ after ``calibration'' studies
at the LHC.

The statistical criterion, equations (1) and (2), apply to the
detected events, i.e., {\em after} efficiency corrections are applied.
Assuming 85\% detection efficiency for a single isolated
lepton,{\cite{sdctdr}} our criterion for the $W^+W^+$ signal
applied to the uncorrected yields is
$\sigma^{\uparrow}>6$ and $\sigma^{\downarrow}>3.5$.

The worst case is at an intermediate value of $m_{\rho}$ for which
neither the $WZ$ nor the $W^+W^+$ signal is large.
In \cite{mcwk1} and in this study we find it is at
$m_{\rho} \simeq 2.5$ TeV, corresponding to minimal technicolor with
$N_{TC}=2$.
The best signal for that case
is in $W^+W^+\ +\ W^-W^-$ scattering. We found in \cite{mcwk1}
that it could be observed
with 63 fb$^{-1}$, which was the basis for our estimate of the ``no-lose''
luminosity.

\noindent {\it \underline {The $\overline qq \rightarrow l\nu
\overline ll$ Background}}

We now consider the background to the $W^+W^+$ signals
from $\overline qq \rightarrow l^+ \nu l^+ l^-$
where the $l^-$ escapes detection. Any detector will have
unavoidable blind spots at low
transverse momentum and at high rapidity. At very low $p_T$
muons will not penetrate the muon detector, electrons or muons may be
lost in minimum bias pile-up, and for low enough $p_T$ in a solenoidal detector
they will curl up unobservably within the beam pipe.
Muon and electron coverage is also not likely to
extend to the extreme forward, high rapidity region.

We have tried to make reasonable though
aggressive assumptions about the observability of the extra electron or
muon.\footnote{We thank K. Einsweiler and M. Gilchriese for several very useful
discussions of the possible capabilities of the ATLAS detector.}
We assume rapidity coverage for electrons and muons of
$\eta(l) < 3$, roughly as expected for the ATLAS
detector. Within this rapidity range we assume that isolated $e^-$ and $\mu^-$
leptons with $p_T(l) > 5$ GeV can be identified in events containing two
isolated, central, high $p_T$ $e^+$'s and/or $\mu^+$'s.
We assume that electrons
(but not muons) with $1< p_T(l) < 5$ GeV can be identified if they are
sufficiently collinear ($m(e^+e^-) < 1$ GeV) with a hard positron in the
central region. For $p_T(e^-) < 1$ GeV we consider electrons to be
unobservable.

It is instructive to examine the $\eta(e^-)$ and $p_{T}(e^-)$ distributions
for electrons that survive this set of cuts. They are shown in figure 1
for $\overline qq \rightarrow e^+e^-\ l^+\nu$ events where $l=e$ or $\mu$.
The two positive leptons are required to satisfy
$|\eta(l^+)| < 2.0$,
$p_T(l^+) > 60$ GeV, and cos$\phi(l^+l^+) < -0.6$. The solid lines are based on
the full set of 10 Feynman diagrams while the dashed lines are for
the $W^+Z$ contribution only, defined as events for which the $l^+ \nu_l$ and
$e^+e^-$ masses lie within $\pm 10$ GeV of $M_W$ and $M_Z$ respectively.
The $WZ$ contribution
dominates for $\eta(e^-) > 3$ and $p_T(e^-) > 5$ GeV, while the other
components dominate for $\eta(e^-) < 2$ and $p_T(e^-) < 5$ GeV.
The total number of events contributing to figure 1 is 27, of which just half,
13.7, are from $WZ$ production.

Another noteworthy feature of the $l^+l^-l^+\nu$ background is the pronounced
peaking of the $l^+$ rapidity distribution toward large $\eta(l^+)$. While the
irreducible backgrounds are also more sharply forward peaked than the
relatively
isotropic signal, the $\eta(l^+)$ distribution from $l^+l^-l^+\nu$ is even more
strongly peaked. In
\cite{mcwk1} we found no advantage from tightening the rapidity cut below
$\eta(l^+)^{MAX} = 2$, but we now find that smaller $\eta(l^+)^{MAX}$
increases the significance of the signal for some models.

A crucial issue beyond the scope of this work is the efficiency with
which the extra lepton can be rejected when it falls within the specified
acceptance region. Following the slogan ``when in doubt, throw it out'',
background veto efficiencies much higher than signal detection efficiencies
should be possible. The effect of an aggressive veto on the
signal efficiency must be considered, but we suspect it is not great. We
have studied all four models assuming 100\% veto efficiency and
have also studied the
effect in the worst-case model of 98, 95 and 90\% veto efficiencies.

\noindent {\it \underline {Results}}

As in \cite{mcwk1} we veto events with central,
high transverse momentum jets ($\eta_J<2.5$ and $p_{TJ}>60$ GeV).
Together with the leptonic cuts this virtually
eliminates the irreducible O($\alpha_W
\alpha_S$) gluon exchange background and is also quite effective
against the O($\alpha_W^2$) background amplitude.\cite{ttbarW}
We scan the three dimensional
parameter space of cuts on the like-sign leptons, consisting
of the single lepton rapidity $\eta^{MAX}(l)$, the single lepton
transverse momentum
$p_T^{MIN}(l)$, and the azimuthal angle between the like-sign leptons,
[cos$\phi(ll)]^{MAX}$.

The signal amplitude is determined by the mass and width of
the $\rho$ meson. We consider minimal one-doublet technicolor with
$N_{TC}=2,\ 3,\ 4$ which by large $N_{TC}$ scaling from QCD implies
$m_{\rho} \simeq 2.52,\ 2.06,\ 1.78$ TeV respectively. To illustrate the
possibility that the mass scale of the new quanta could be even heavier,
we consider a
fourth model with $m_{\rho}$ set to an arbitrarily large value, 4 TeV,
and the width $\Gamma_{\rho}$ fixed by scaling from QCD.

Figure 2 shows the integrated, single lepton transverse momentum
distribution for $l^+l^+ + l^-l^-$ leptons, i.e., the number of events
per 100 fb$^{-1}$ with like-sign lepton
$p_T(l) > p_T^{MIN}(l)$ as a function of $p_T^{MIN}(l)$. The signal is for
$m_{\rho}=2.52$ TeV, and $\eta^{MAX}(l)$ and [cos$\phi(ll)]^{MAX}$ are
set to their optimum values (see table 1). At the optimum,
$p_T^{MIN}(l)=130$ GeV, the signal is $\sim 2.5$ times
bigger than the background, with equal contributions
from $\overline qq \rightarrow l^{\pm}\nu_ll^+l^-$
and the irreducible background.

For each model and
for every point in the three dimensional parameter space ($\eta^{MAX}(l),\
p_T^{MIN}(l),\ {\rm [cos}\phi(ll)]^{MAX}$) we computed the integrated
luminosity needed to satisfy the acceptance corrected observability criterion.
The smallest is then the minimum luminosity for that model,
${\cal L}_{MIN}$, and the corresponding point in the ($\eta^{MAX}(l),\
p_T^{MIN}(l),\ {\rm [cos}\phi(ll)]^{MAX}$) parameter space is the
optimal cut. Table 1 reports the minimum luminosity for each of
the four values of
$m_{\rho}$ for both the resonant $WZ$ and nonresonant like-sign $WW$ signals.
In each case both charge channels are included,
that is, $W^+Z\ +\ W^-Z$ and $W^+W^+\ +\ W^-W^-$.

For the resonant $WZ$ channel table 1 records only ${\cal L}_{MIN}(WZ)$
--- details can be found in \cite{mcwk1}.
For the like-sign $WW$ signals the table includes the minimum
luminosity, the
optimized cuts on the like-sign leptons, the resulting number of signal and
background events per 100 fb$^{-1}$, and the relative contributions of the
three
background components.

Complementarity of the resonant and nonresonant channels is evident in the
inverse relationship between ${\cal L}_{MIN}(WZ)$ and ${\cal L}_{MIN}(WW)$
in the table. For the like-sign $WW$ channel the
biggest signal occurs for the heaviest $\rho$
meson, $m_{\rho}=4$ TeV, with a signal meeting the significance criterion for
77 fb$^{-1}$. The $WZ$ signal is smallest for that case, and there is
no cut for which equation 3 is satisfied, indicated by ``NS'' (no signal) in
the table. For $m_{\rho}=1.78$ TeV the
like-sign $WW$ signal is smaller but the resonant $WZ$ signal is much larger
and
satisfies the
criterion with only 44 fb$^{-1}$. The worst case among the four models
is at $m_{\rho}=2.52$ TeV, as in our previous study. Because of the
$l^+l^-l^+\nu$ background we now find that the
optimum cut for that model, still in the like-sign $WW$ channel, requires
105 fb$^{-1}$ to meet the criterion, compared to 63 fb$^{-1}$ in \cite{mcwk1}.

In table 1 we assumed 100\% veto efficiency when the
third lepton falls within the specified geometric acceptance.
In table 2 we study the effect of veto inefficiency
for $m_{\rho}=2.52$ TeV. At 98\% efficiency the effect
is not great but at 95\% ${\cal L}_{MIN}(WZ)$ is increased by 40\%. For 90\%,
not shown in the table, ${\cal L}_{MIN}(WZ)$ would be nearly doubled, to
200 fb$^{-1}$. We presume that an aggressive $\simeq 98\%$ efficient veto is
possible without significantly affecting the signal efficiency, but its
feasibility requires further study.

\noindent{\it \underline {Sensitivity to Collider Energy and Detector
Coverage}}

The relative importance of the $\overline qq \rightarrow
l^+ \nu l^+ l^-$ background to the $W^+W^+$
signal depends sensitively on the collider energy.
At $\sqrt{s}=14$ TeV
it doubles the background for the optimum cuts and increases
${\cal L}_{MIN}$ for $m_{\rho}=2.52$ TeV by 50\%, from 63 fb$^{-1}$
to 105 fb$^{-1}$. At higher energy the effect is smaller.
For instance, for $\sqrt{s}=40$ TeV and
$m_{\rho}=2.52$ TeV the  $\overline qq \rightarrow
l^+ \nu l^+ l^-$ process increases the background by 40\% and ${\cal L}_{MIN}$
by 25\%, from 5.2 fb$^{-1}$ to 6.5 fb$^{-1}$. Since $\overline qq
\rightarrow l^+ \nu l^+ l^-$ is dominated by $\overline qq
\rightarrow WZ$ and $\overline qq \rightarrow W\gamma^*$, this energy
dependence
follows from their essentially two body phase space
compared to the four body phase space of the
$qqW^+W^+$ signal and irreducible background.

We have made a preliminary exploration of the sensitivity to the
acceptance region for the third lepton, with some unexpected results.
For the results quoted above we assumed that wrong-sign electrons or
muons could be detected for $\eta < 3$ and $p_T > 5$ GeV. If the coverage for
the wrong-sign lepton is increased ambitiously to $\eta < 5$ the gain
is surprisingly little. For $m_{\rho}=2.52$ TeV with
the cuts specified in table 1 the total background decreases by just
$\sim 10$\% and ${\cal L}_{MIN}$ only goes from 105 to 101 fb$^{-1}$.
Reoptimizing the cuts
we find ${\cal L}_{MIN} = 98$ fb$^{-1}$, a negligible change. This
reflects the importance of the low $p_T$ leptons.

On the other hand if we relax the transverse momentum cut for the wrong-sign
lepton veto from 5 to 10 GeV, the effect is substantial. The result for
$m_{\rho}=2.52$ TeV with the table 1 cuts is to increase the
$l^+ \nu l^+ l^-$ background by a factor $\sim 2.5$. Reoptimizing the cuts we
now find that ${\cal L}_{MIN}$ would increase by 30\% to 138 fb$^{-1}$.

\noindent{\it \underline {$qq \rightarrow qqWZ$ Background to $qq \rightarrow
qqW^+W^+$ Signal?}}

In our study of the $WZ$ channel
in \cite{mcwk1} we found important contributions to both signal and background
from $\overline qq$ annihilation ($\overline qq \rightarrow \rho \rightarrow
WZ$ and $\overline qq \rightarrow WZ$) and from $qq \rightarrow qqWZ$.
Because of the energy dependence of
two and four body phase space, we found that
$\overline qq$ annihilation dominates at the LHC while $qq \rightarrow qqWZ$
dominates (or would have) at the SSC. However even at the LHC the $qq
\rightarrow qqWZ$ process is not negligible, contributing of order 30\% of the
signal and background. Since in this paper we have established a large
background to $W^+W^+$ scattering from $\overline qq \rightarrow W^+Z$, we
should also consider the possible background from $qq \rightarrow qqW^+Z$.

We have done so and find that it is very small. With $m_H \leq 100$ GeV
and assuming that the
wrong-sign lepton is unobservable for $\eta > 3$ or for $p_T < 5$ GeV, we
find only 0.041 events per 100 fb$^{-1}$ from $qq \rightarrow
qq W^{\pm}Z$ for the cuts that optimize
the $W^+W^+$ signal for $m_{\rho}=2.52$ TeV
(see table 1). We have
not computed the complete $qq \rightarrow qql^{\pm}\nu l^+l^-$ amplitude
since it
would not be important even if it were a few times larger than the
$qq \rightarrow qqW^{\pm}Z$ component.

\noindent{\it \underline {Discussion}}

In evaluating the $qq \rightarrow qqWZ$ background to $W^+W^+$ scattering we
considered the order $\alpha^2_W$ standard model amplitude with $m_H \leq
100$ GeV. Just as for $\overline qq \rightarrow WZ$, we did not consider the
enhanced $WZ$ cross section that would occur if a $\rho$ resonance were within
the range of the LHC. That is because the precise question we are
addressing is whether the standard model,
with no strong force in the symmetry breaking sector,
can contribute backgrounds that would be confused with
strong $WW$ scattering. Enhanced $WZ$ production from strong dynamics
in the $WZ$ channel would indeed increase the apparent $W^+W^+$ signal,
and the true $W^+W^+$ cross section could only be disentangled after
measuring $\sigma(WZ)$. But since
the overriding question initially is simply
whether there is or is not strong dynamics in the symmetry breaking sector,
this false amplification of the $W^+W^+$ cross section is actually
advantageous:
though it exaggerates the size of $\sigma(W^+W^+)$ it {\em is} truly an
effect of strong symmetry breaking dynamics. As such it tends to offset the
suppression of the signal by the $l^+\nu l^+l^-$ background.
We will report elsewhere on a study of this pseudo-amplification of the
$W^+W^+$ cross section as a function of $m_{\rho}$.

In our previous study we claimed, extrapolating from studies of the QCD
corrections to $\overline qq \rightarrow ZZ$, that the lowest order
amplitude gives a conservative estimate of
the $\overline qq \rightarrow WZ$ background provided we use
a central jet veto.
This conjecture is now supported by a study of the next to leading order (NLO)
cross section for $WZ$ production.\cite{wzqcd} As for $ZZ$ production, the
$qg \rightarrow qWZ$ process greatly enhances the single boson ($W$ or $Z$)
spectrum at high $p_T$, but the
enhancement is associated with a momentum-balancing high $p_T$ quark jet
collinear with the second weak boson.
It is therefore not relevant to our study because we use a central jet veto,
which typically suppresses the single boson high-$p_T$ spectrum to
be at or just below the level of the tree approximation.\cite{wzqcd}
It would be
useful to compute the NLO cross section for the cuts used here.

We are surprised by our own conclusion that $\overline qq \rightarrow
l^+ \nu l^+ l^-$ is one of the two dominant backgrounds to strong $W^+W^+$
scattering assuming 100\% veto efficiency for the third lepton, and is the
dominant background for veto efficiency $\leq 98\%$, as shown in table 2.
Despite the sizeable increase in our estimate of
the background, we find that the
LHC operating at 14 TeV and 10$^{34}$ cm.$^{-2}$ sec.$^{-1}$ would be able in
$\sim 1$ to $\sim 1 {1\over 2}$ years to provide a significant strong
scattering
signal in the worst case scenario. The prospects might be better than we
have indicated because of the ``pseudo-amplification'' of the $W^+W^+$ signal
by enhanced strong scattering in the $W^+Z$ channel, which remains to be
studied
in detail. A critical experimental question requiring further study is the
efficiency with which the extra leptons can be vetoed when they fall within the
acceptance of the detector.

\vskip .2in
\noindent Acknowledgements: We are grateful to G. Azuelos, K. Einsweiler,
M. Gilchriese and H. Murayama
for helpful discussions and suggestions. This work was supported
by the Director, Office of Energy
Research, Office of High Energy and Nuclear Physics, Division of High
Energy Physics of the U.S. Department of Energy under Contracts
DE-AC03-76SF00098 and DE-AC02-76CHO3000.

\newpage

Table 1. Minimum luminosity to satisfy significance criterion for $W^+Z + W^-Z$
and $W^+W^+ + W^-W^-$ scattering. For the like-sign $WW$ channel we also
specify
the optimum cut on the like-sign leptons that gives ${\cal L}_{MIN}(WW)$,
the corresponding number of signal and background events per 100 fb$^{-1}$, and
the compositon of the background for
the optimum cut. A central jet veto is included as specified in the text.
Rejection of all events for which the third lepton
falls within its acceptance region is assumed.

\begin{center}
\vskip 20pt
\begin{tabular}{ccccc}
$m_{\rho}$(TeV) & 1.78 & 2.06&2.52&4.0\cr
\hline
\hline
&&&&\cr
${\cal{L}}_{MIN}(WZ)\ ({\rm{fb}}^{-1})$ &44 &98 &323&NS\cr
${\cal{L}}_{MIN}(WW)\ ({\rm{fb}}^{-1})$&142&123&105&77\cr
&&&&\cr
\hline
&&&&\cr
$WW$ Cut&&&&\cr
&&&&\cr
$\eta^{MAX}(l)$&1.5&1.5&1.5&2.0\cr
$p^{MIN}_{T}(l)$ (GeV)&130.&130.&130.&130.\cr
$[\cos\phi(ll)]^{MAX}$&$-0.72$&$-0.80$&$-0.80$&$-0.90$\cr
&&&&\cr
\hline
&&&&\cr
$WW$ Sig/Bkgd&&&&\cr
&&&&\cr
(events per $100\ {\rm{fb}}^{-1})$&12.7/6.0&14.1/5.8&15.9/5.8&22.4/8.9\cr
&&&&\cr
\hline
&&&&\cr
$WW$ Backgrounds (\%)&&&&\cr
&&&&\cr
$\overline l l \overline l \nu_{l}$&47&49&49&61\cr
$O(\alpha^2_W)$&47&46&46&33\cr
$O(\alpha_W\alpha_S)$&6&6&6&6\cr

\end{tabular}
\end{center}

\newpage
Table 2. Minimum luminosity to satisfy significance criterion for
$W^+W^+ + W^-W^-$ scattering for $m_{\rho}=2.52$ TeV,
assuming 100\%, 98\% or 95\% efficiency for the
veto of wrong-sign charged leptons that fall
within the acceptance region specified in
the text. The optimum cuts and corresponding yields are
shown as in table 1.

\begin{center}
\vskip 20pt
\begin{tabular}{ccccc}
Efficiency & 100\% & 98\% & 95\% \cr
\hline
\hline
&&&&\cr
${\cal{L}}_{MIN}(WW)\ ({\rm{fb}}^{-1})$&105&115&148 \cr
&&&&\cr
\hline
&&&&\cr
$WW$ Cut&&&&\cr
&&&&\cr
$\eta^{MAX}(l)$&1.5&1.5&1.5\cr
$p^{MIN}_{T}(l)$ (GeV)&130.&130.&160.\cr
$[\cos\phi(ll)]^{MAX}$&$-0.80$&$-0.80$&$-0.86$\cr
&&&&\cr
\hline
&&&&\cr
$WW$ Sig/Bkgd&&&&\cr
&&&&\cr
(events per $100\ {\rm{fb}}^{-1})$&15.9/5.8&15.9/7.9&12.1/5.6\cr
&&&&\cr
\hline
&&&&\cr
$WW$ Backgrounds (\%)&&&&\cr
&&&&\cr
$\overline l l \overline l \nu_{l}$&49&62&71\cr
$O(\alpha^2_W)$&46&34&26\cr
$O(\alpha_W\alpha_S)$&6&4&3\cr

\end{tabular}
\end{center}

\newpage
\begin{center}
{\bf Figure Captions}
\end{center}
\vskip .5 in
\noindent Figure 1. The $e^-$ rapidity (figure 1a) and transverse momentum
(figure 1b) distributions corresponding to 100 fb$^{-1}$
for $\overline qq \rightarrow e^+e^-\l^+\nu_{l}$ where $l=e$ or $\mu$. The cuts
on the electron and the positive leptons are specified in the text, and
100\% rejection is assumed for electrons within the geometric
region specified for the veto.
The dashed lines show the contribution from $\overline qq \rightarrow W^+Z$.

\vskip .15in
\noindent Figure 2. The number of events per 100 fb$^{-1}$
for which both like-sign leptons have
transverse momentum greater than $p_T^{MIN}$. The rapidity and azimuthal angle
cuts on the like-sign leptons are at the optimum values specified in table 1
for $m_{\rho}=2.52$ TeV. All events with the third lepton
inside its acceptance region are rejected.
The solid, dashed, dot-dashed, and dotted lines are respectively
the signal and the backgrounds from $\overline qq \rightarrow l^{\pm} \nu_l
\overline ll$ and from
$qq \rightarrow qqW^+W^+/W^-W^-$ in orders $\alpha_W^2$ and $\alpha_W
\alpha_S$.


\begin{thebibliography}{99}
\bibitem{strongww} M.S. Chanowitz and M.K. Gaillard, {\it Nucl. Phys.} B261,
379
(1985).
\bibitem{mcwk1} M.S. Chanowitz and W. Kilgore, {\it Phys. Lett.} B322, 147
(1994).
\bibitem{baggeretal} J. Bagger et al., {\it Phys. Rev.} D49, 1246 (1994).
\bibitem{ttbar} D. Dicus, J. Gunion, and R. Vega, {\it Phys. Lett.} 258B,
475 (1991); D. Dicus et al., {\it Nucl.Phys.} B377, 31 (1992).
\bibitem{ttbarW} V. Barger et al., {\it Phys. Rev. } D42, 3052 (1990).
\bibitem{azuelosetal} G. Azuelos, C. Leroy, and R. Tafirout, ATLAS Internal
Note
PHYS-NO-033, 1993 (unpublished); G. Azuelos, talk presented at CERN,
April 21, 1994 (unpublished), and update to appear.
\bibitem{helas} H. Murayama, I. Watanabe, and K. Hagiwara, KEK Report 91-11,
1992.
\bibitem{madgraph} T. Stelzer and W.F. Long, {\it Comp. Phys. Commun.} 81, 357
(1994).
\bibitem{rhomodel} S. Weinberg, {\it Phys. Rev.} 166, 1568 (1968);
R. Casalbuoni et al., {\it Phys. Lett.} 155B, 95 (1985).
\bibitem{sdctdr} Solenoidal Detector Collaboration, E.L. Berger et al.,
{\it Technical Design Report}, SDC-92-201, 1992.
\bibitem{wzqcd} U. Baur, T. Han, and J. Ohnemus, FSU-HEP-941010, 1994
(submitted
for publication).

\end{thebibliography}
\end{document}